\documentstyle[prb,aps,epsf]{revtex}

\begin{document}
\draft

\preprint{UdeM.\ Rep.\ No.\ PMC/LJL/97-xx}

\twocolumn[\hsize\textwidth\columnwidth\hsize\csname @twocolumnfalse\endcsname

\title{\bf Melting, freezing, and coalescence of gold nanoclusters}
\author{Laurent J.\ Lewis,\cite{byline1} Pablo Jensen, and Jean-Louis Barrat}
\address{D\'epartement de Physique des Mat\'eriaux, Universit\'e Claude-Bernard
Lyon-I, CNRS UMR 5586, 69622 Villeurbanne C\'edex, France }

\date{\today}

\maketitle

\begin{center}
Submitted for publication in Physical Review B \\
\end{center}

\begin{abstract}
We present a detailed molecular-dynamics study of the melting, freezing, and
coalescence of gold nanoclusters within the framework of the embedded-atom
method. Concerning melting, we find the process to first affect the surface
(``premelting''), then to proceed inwards. The curve for the melting
temperature vs cluster size is found to agree reasonably well with
predictions of phenomenological models based on macroscopic concepts, in
spite of the fact that the clusters exhibit polymorphism and structural
transitions. Upon quenching, we observe a large hysterisis of the transition
temperature, consistent with recent experiments on lead. In contrast, we find
macroscopic sintering theories to be totally unable to describe the
coalescing behaviour of two small clusters. We attribute this failure to the
fact that the nanocrystals are facetted, while the sintering theories are
formulated for macroscopically smooth crystallites. The time for coalescence
from our calculations is predicted to be much longer than expected from the
macroscopic theory. This has important consequences for the morphology of
cluster-assembled materials.
\end{abstract}

\vskip2pc]

\narrowtext

\section{Introduction}\label{intro}

Nanometer-size particles, or clusters, have received much attention
recently.\cite{melinon} From a fundamental point of view, it is of interest
to understand how the properties of nanoparticles are affected by their size,
and in particular how the behaviour of the bulk material is approached. The
field has received considerable impetus from speculations about possible
technological applications of nanoparticles in optoelectronic devices.
Computer simulation methods, and in particular molecular dynamics (MD), have
become a favourite tool for investigating theoretically the physics of
nanoparticles. Indeed, with present-day computers, detailed simulations of
clusters containing several thousands of atoms are easily feasible using
empirical potential models.\cite{simul} More accurate simulations using {\em
ab-initio} interactions are possible for particles containing up to a few
tens atoms and have indeed been used to study small semiconductor
clusters.\cite{abinitio}

In this paper, we use MD simulations to address two issues. First, we
investigate the influence of size on the melting and freezing of gold
nanoparticles. Second, we examine the coalescence of two gold nanoparticles,
both of identical size and not. We focus here on gold because it has been the
object of several experimental studies,\cite{fluelli,borel} and also because
semi-empirical, many-body potentials are available for this material. The
first question has already been addressed by several authors using a variety
of approaches, experimental and
theoretical.\cite{borel,duxbury,tosatti,labastie,celestini} We take it here
as a preliminary step towards the second stage of our study, namely
coalescence. Understanding coalescence is of primary importance for
understanding the structure of cluster-assembled materials. These materials
can be grown by the low-energy deposition of preformed clusters containing
hundreds or thousands of atoms on a surface.\cite{LECD} The structure of the
resulting films depends critically on the diffusion properties of the
deposited clusters,\cite{PJLB,Pierre} as well as on their sintering
properties. On the basis of thermodynamics, it is evident that clusters
deposited on a surface will tend to coalesce and form larger drops. The
kinetics of this process, which is determinant for understanding the actual
structure of the film, is however not known in detail.

Concerning melting, we find the clusters to exhibit premelting prior to the
transition. The melting temperature vs size curve is found to agree
reasonably well with predictions of phenomenological models based on
macroscopic concepts, despite the fact that the clusters exhibit a rather
complex structure --- polymorphism and structural transitions. Upon
quenching, we observe a large hysterisis of the transition temperature. In
contrast, we find macroscopic sintering theories to be totally unable to
describe the coalescing behaviour of two small clusters. We attribute this
failure to the fact that the nanocrystals are facetted, while the sintering
theories assume macroscopically smooth crystallites. We predict the time for
coalescence to be much longer than expected from the macroscopic theories.
This will have important consequences for the morphology of cluster-assembled
materials. We consider here the coalescence of unsupported clusters, i.e., in
vacuum rather than on a substrate. Evidently, an important role of the
substrate in the actual coalescence of supported clusters is to ensure
thermalisation. In the constant-temperature MD simulations reported below,
thermalisation is taken care of by coupling the system to a
``thermostat''.\cite{MD} We therefore expect the coalescence events studied
here to be relevant to supported clusters in the case where they are loosely
bound to the substrate, e.g., gold clusters on a graphite substrate.

The paper is organized as follows. The computational details are described in
Sec.\ \ref{comp_det}. The melting/freezing transition is described and
analyzed in Sec.\ \ref{melt_freeze}. Coalescence of two unsupported clusters
is decribed in Sec.\ \ref{coalesc}, and Sec.\ \ref{concl} summarizes our main
conclusions.

\section{Computational details}\label{comp_det}

\subsection{MD simulations}

In order to be able to simulate clusters containing more than several hundred
particles, as noted above, it is necessary to resort to an empirical
description of the interatomic forces. Here we chose to employ the
embedded-atom method (EAM),\cite{fbd} an $n$-body potential with proven
ability to model reliably various static and dynamic properties of transition
and noble metals, in either bulk or surface configurations.\cite{eam-review}
The model is ``semi-empirical'' in the sense that it approaches the
total-energy problem from a local electron density viewpoint but using a
functional form with parameters obtained from experiment (equilibrium lattice
constant, sublimation energy, bulk modulus, elastic constants, etc.).

In the case of gold, EAM gives an excellent description of jump diffusion on
the (100) surface (the activation energy is predicted to be 0.64eV compared
to 0.62 eV from first principles),\cite{blpn95} but does not do as well on
the (111) surface, where the barrier is very low (0.02 vs 0.20
eV).\cite{blpn95} The melting temperature for bulk Au is predicted by EAM to
be 1090 K. This does not agree well with the experimental value of 1338 K.
Thus, while we expect the model to give a qualitatively correct description
of the nanoclusters, we also expect significant differences to show up in the
numerical values of the calculated properties. One interesting advantage of
EAM (and similar models) vs more generic models such as Lennard-Jones,
however, is that the $n$-body formalism gives a much better description of
cohesion in non-equilibrium situations, as is the case on a surface at finite
temperature. Thus, for instance, while desorption of atoms on a Lennard-Jones
surface is unphysically large, it occurs very rarely on a EAM surface. Hence,
in spite of the quantitative limitations of EAM, we expect the model to
provide a qualitatively correct description of the system.

The MD calculations were performed using a parallel version of program {\tt
groF}, a general-purpose MD code for bulk and surfaces developed by one of
the authors (LJL), optimized to run on the Convex of the PSMN/\'ENS.\cite{PSMN}
The program employs a predictor-corrector algorithm to integrate the
equations of motion, and provides the option to do extended-system
simulations (constant temperature and/or constant pressure). In all the
simulations reported here, we used a timestep of 2.5 fs; this is a fairly
small value, which we judged necessary in order to ensure proper stability of
the trajectories during the very long runs needed to study coalescence (up to
10 ns, i.e., 4 million steps).

\subsection{Cluster preparation}

\begin{figure}
\epsfxsize=10cm
\epsfbox{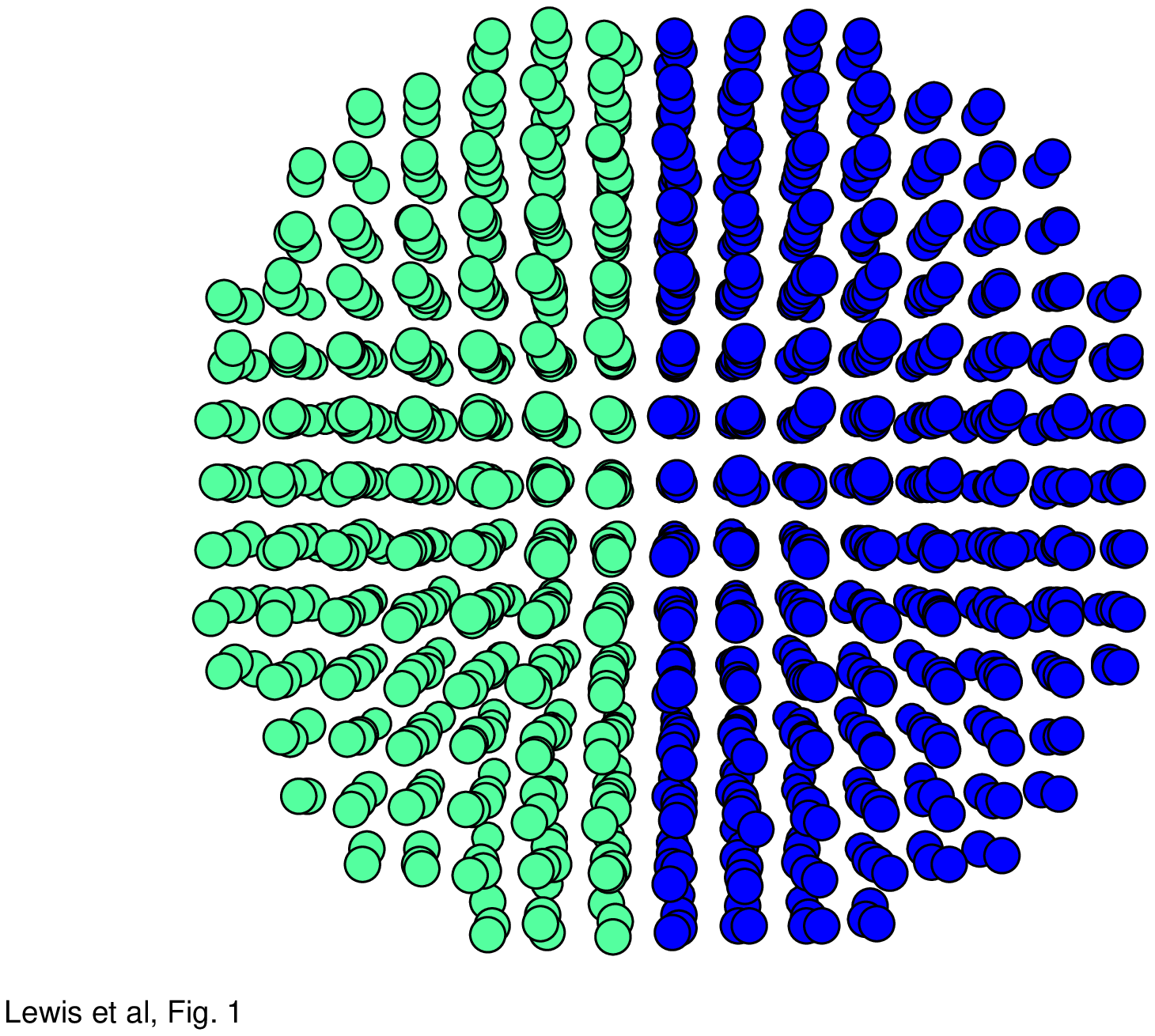}
\vspace*{-5cm}
\caption{
Structure of the 1055-atom Au cluster in the fcc structure, i.e., before
melting and re-solidifying, at 300 K. The two halves of the cluster are
represented with different colors in order to facilitate visualisation of
diffusion processes.
\label{1055_300}
}
\end{figure}

The clusters used to initiate the melt-and-freeze runs were prepared as
follows: Starting with a large block of face-centered cubic Au, ``spherical''
clusters were carved out such that the center-of-mass coincides exactly with
an atom. Of course, the clusters cannot be perfectly spherical and will, even
in their ground state, exhibit facets. A 1055-atom cluster is shown in Fig.\
\ref{1055_300}. (In order to better follow diffusion processes, the two
halves of the cluster are colored differently.) We did not consider using
other cluster geometries, such as icosahedral, octahedral, or Wulff
polyhedral.

Clusters of eight different sizes were considered, from 135 to 3997 atoms, as
indicated in Table \ref{clusters}, thus spanning a fairly wide range of
melting temperatures (see Section \ref{melt_freeze}). It should be noted that
small clusters can undergo several structural transitions as a function of
temperature (below melting), and therefore constitute an object of study in
their own right. We actually observed many such transitions in our smaller
clusters, but this is beyond the scope of the present study.

After equilibration at low temperature, the clusters were subjected to a
melt-and-quench cycle in order to identify the melting temperature and to
force the clusters into their ground state. (For the reasons mentioned above
only the larger clusters --- 767 atoms and beyond --- were re-solidified
after melting). Upon approaching the transition from either side, the
temperature was changed in steps of 25 to 100 K. At each temperature, the
system was first fully equilibrated before running to accumulate statistics.
Typically, runs ranged from 100 to 250 ps, depending on the ``distance'' from
the transition, which could be anticipated from the behaviour of the
potential energy. The melt-and-freeze simulations were carried out in the
microcanonical ensemble.

\section{Melting and freezing}\label{melt_freeze}

\subsection{Total energies}

There are several ways of proceeding for identifying the melting and freezing
transitions. The simplest is perhaps to examine the variations of potential
energy with temperature. This is done in Fig.\ \ref{E_vs_T}(a) for the
1055-atom cluster. The ``as-made'' fcc cluster was first equilibrated at 300
K, then heated up slowly until it melted. The melting transition is clearly
identifiable by the large upward jump in energy at a temperature of about 835
K; on either side of the transition, the energy varies smoothly, almost
linearly with temperature. Very near the transition, the system becomes
unstable and the data points are characteristic of a transient state. Upon
cooling from the highest temperature, the system undergoes a sharp
liquid-solid transition in spite of a rather strong hysterisis. Clearly, the
new solid phase, as far as energy is concerned, is equivalent to the initial
one, though, as we will see below, there are some structural differences.

It is much easier for a cluster to go from an ordered state to a disordered
state than the opposite, i.e., in the present context, to melt than to freeze
during the finite time covered during the simulations. This explains, in
part, the hysteresis we observe in the melt-and-quench cycle, and also
indicates that the melting temperature is probably much closer to the
thermodynamic transition point than the freezing temperature. However,
hysteresis in the melting/freezing transition is expected theoretically,
\cite{reiss} and has also been reported experimentally in the case of
lead.\cite{kofman}

In Fig.\ \ref{E_vs_T}(b) we show the potential energy for the 767-atom
cluster. This system exhibits interesting behaviour. First, upon heating
(solid line), we see that the cluster undergoes a solid-solid transition (of
the type mentioned above) to a lower-energy phase (which we have not analyzed
in detail), at a temperature of about 710 K, then melts at about 775 K. Upon
cooling, freezing takes place at about 680 K, but now to a state that lies
somewhat lower in energy than the initial phase --- likely the same as that
which appeared just below melting as can be inferred from their comparable
energies.

The structure of the low-temperature, re-solidified clusters is discussed in
more detail below; they differ from the initial, low-temperature, fcc
clusters in that large facets are now present. Because of finite-time
limitations in the MD simulations, the transition to the ground-state
structure is however incomplete: while facets develop, the melt-and-quenched
clusters are ``packed'' with defects so that, in effect, their energies lie
above those of the corresponding fcc clusters for large enough sizes (1505
and beyond).

\subsection{Dynamics and density distributions}

Melting and freezing of the clusters can also be quantified in terms of
diffusion, as is often done for the corresponding bulk transitions. It is
evidently most appropriate, in the case of a cluster, to examine diffusion as
a function of distance from the surface. Thus, we may define radial bins
about the center of mass and calculate, for an average atom within each bin,
the mean-square displacement.

\begin{figure}
\vspace*{1cm}
\epsfxsize=7cm
\epsfbox{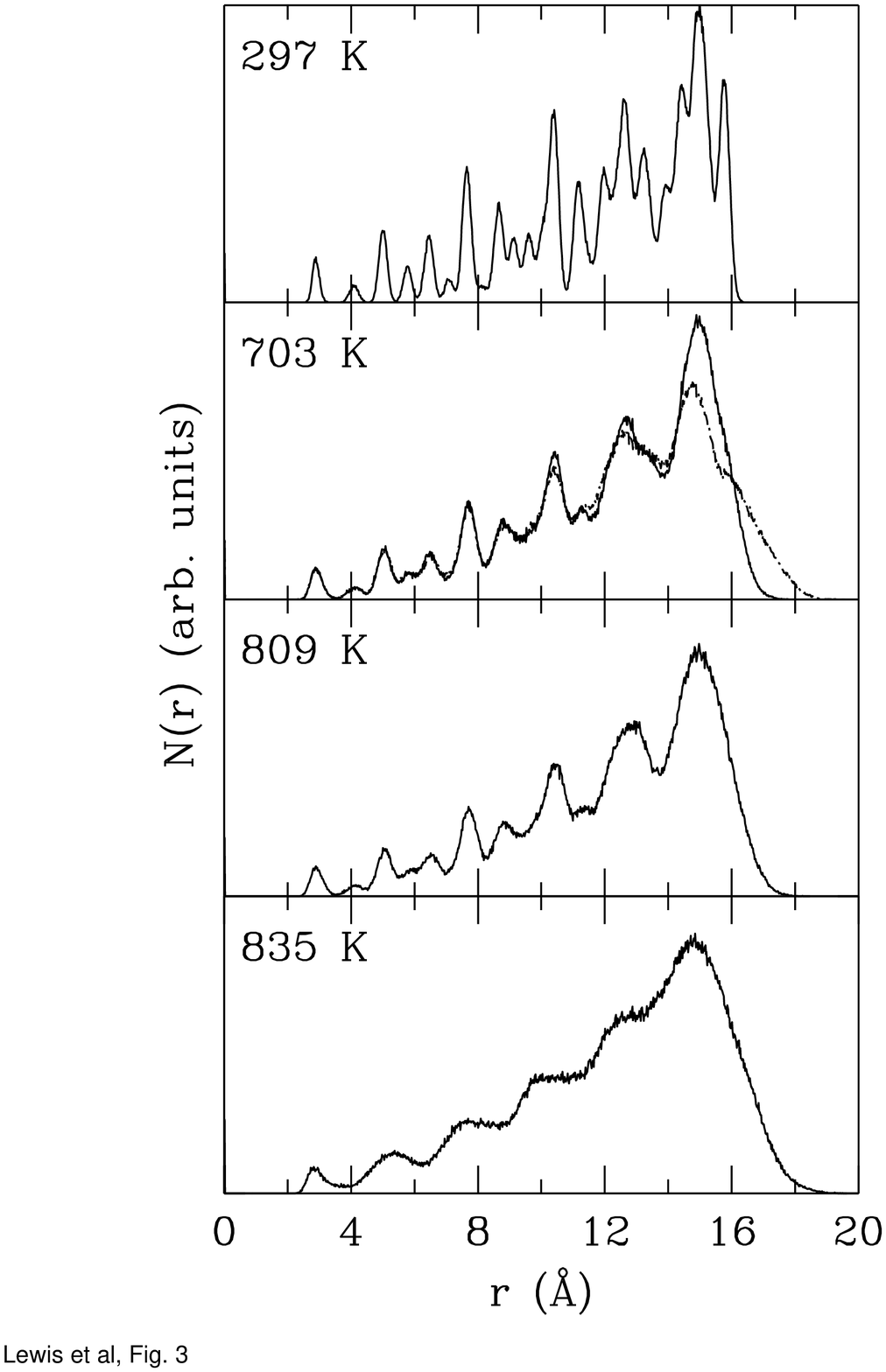}
\vspace*{2cm}
\caption{
Density distribution $N(r)$ for the 1055-atom cluster at four different
temperatures, as indicated. At $T=703$ K, we show the results for both the
initial fcc cluster (full line) and the re-solidified cluster (701 K, dashed
line). The cluster at 835 K, evidently, is molten.
\label{nofr}
}
\end{figure}

In order to define those bins in a physically-meaningful manner, we introduce
the density distribution function $N(r)$, where $N(r)dr$ is the number of
atoms within a shell of thickness $dr$ at $r$ from the atom at the center of
mass of the cluster. We show this quantity at 5 different temperatures for
the 1055-atom cluster in Fig.\ \ref{nofr}: 297, 703, 809, 835, upon heating,
and 701 upon cooling, averaged over thousands of independent configurations.
The results of runs 703 (heating) and 701 (cooling) are superimposed on the
same graph so as to display the similarities and differences between the two
structures.

From Fig.\ \ref{nofr} we see that $N(r)$ for the solid phases of the cluster
displays a structure typical of a crystal at finite temperature, i.e., an
elaborate series of well-defined peaks broadened by thermal agitation. This
is of course particularly true of the inner shells, where the structure is
more bulk-like. The outer shells, in contrast, feel strongly the influence of
the surface, even more so that the temperature is high. This is in fact very
reminiscent of corresponding situation in the case of infinite surfaces.

Upon melting, just like a bulk crystal does, $N(r)$ exhibits a behaviour
similar to that of an ordinary, bulk liquid, i.e., a series of peaks that
merges into a featureless continuum at large distances (835 K in Fig.\
\ref{nofr}). In particular, small peaks merge and the minimum between the
first- and second-neighbour peak is filled up, indicating that diffusion is
actively taking place.

\begin{figure}
\vspace*{1cm}
\epsfxsize=7cm
\epsfbox{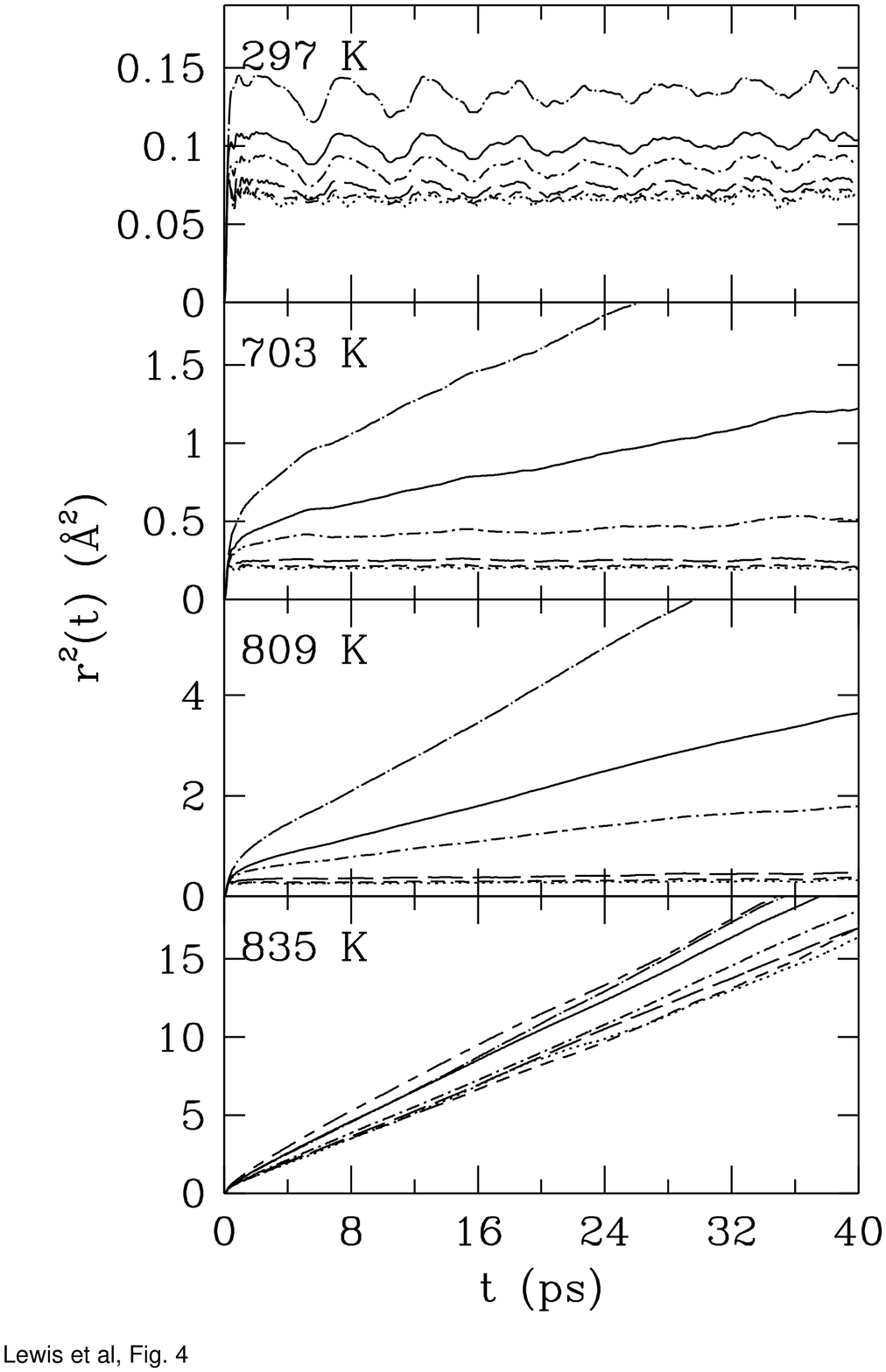}
\vspace*{2cm}
\caption{
Mean-square displacements for the 1055-atom cluster at four different
temperatures, as indicated, for a series of concentric shells as discussed in
the text; the full line, at each temperature, corresponds to the average
(over all particles) mean-square displacement.
\label{msd}
}
\end{figure}

It is easy to see from Fig.\ \ref{nofr} at 835 K that the liquid cluster
consists of concentric shells of atoms of thickness roughly equal to the
hard-sphere diameter, i.e., about 2.2--2.4 \AA\ here: the atoms are
``rolling'' on top of one another in such a way that the inherent packing
arrangement dictated by the central atom is preserved. It is therefore
natural to examine diffusion as a function of the radial position of the
shell. This is done in Fig.\ \ref{msd}, where we plot the mean-square
displacements $r^2(t)$ versus time for the same four temperatures as in Fig.\
\ref{nofr}. Of course, inter-shell diffusion is also taking place, so in fact
an atom will sample a region of radial space that extends beyond its own bin.
Here, for a given bin, we average over all atoms that belonged to that bin at
$t=0$. In the high-$T$ limit, for long enough averaging time, all particles
will sample evenly the whole cluster.

The mean-square displacements at very low temperatures exhibit a behaviour
which is typical of a cold crystal with a surface: All shells possess a
vanishing diffusion constant (as far as we can tell on the timescale covered
by the simulations, viz.\ $\sim$100 ps) and the amplitude of the oscillations
of the atoms about their equilibrium positions increases upon going from the
core to the surface, here by a factor of roughly two.

At 703 K, now, we see that surface diffusion is taking place (this was
already visible at somewhat lower temperatures not shown in Fig.\ \ref{msd}),
but affects only the outer shell. In fact, the corresponding distribution
function reveals that the outer shell peak has lost most of its crystalline
character and possesses a rather well-defined liquid-like structure. At 809
K, $N(r)$ is even more liquid-like at the surface, and in fact the two outer
shells now possess non-zero diffusion constants. The core of the cluster
however remains crystalline. This behaviour, of course, is related to surface
``premelting'', as observed in the case of large crystals--- see, e.g.,
Refs.\ \onlinecite{tosatti} and \onlinecite{zangwill}. At the next
temperature shown, finally, 835 K, i.e., right above the melting transition,
the system is completely liquid, as can be inferred from the shape of $N(r)$
but also from the non-zero, large, and almost identical, diffusion constants
for all shells.

\subsection{Atomic structure}

\begin{figure}
\epsfxsize=10cm
\epsfbox{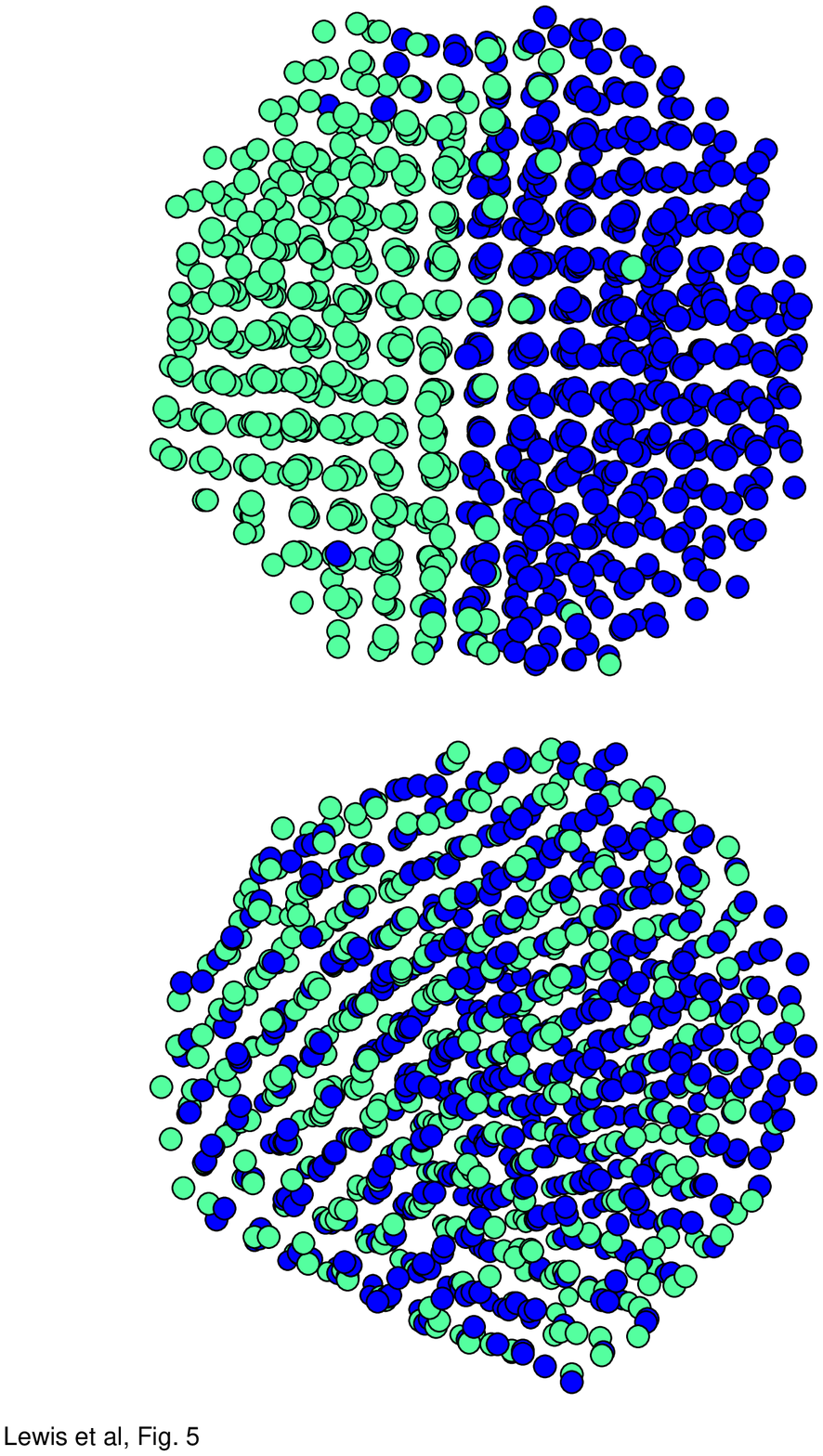}
\vspace*{-1cm}
\caption{
Structure of the 1055-atom cluster (a) at 809 K, before melting (i.e., in the
fcc structure) and (b) at 805 K, after melting and re-solidifying.
\label{1055_800}
}
\end{figure}

An interesting characteristic of the hot-solid phase of the cluster is that
even though the {\em surface} is definitely liquid --- cf.\ Fig.\ \ref{msd}
at 809 K --- it exhibits very well-defined facets, i.e., the cluster is {\em
not} spherical. This can be seen clearly in Fig.\ \ref{1055_800}(a) for the
1055-atom cluster at 809 K --- about 25 K below melting --- and even more so
in Fig.\ \ref{1055_800}(b) at the same temperature, but after melting and
freezing. Such facets reflect the surface anisotropy induced by the core of
the cluster, which remains solid at this temperature.

\begin{figure}
\epsfxsize=10cm
\epsfbox{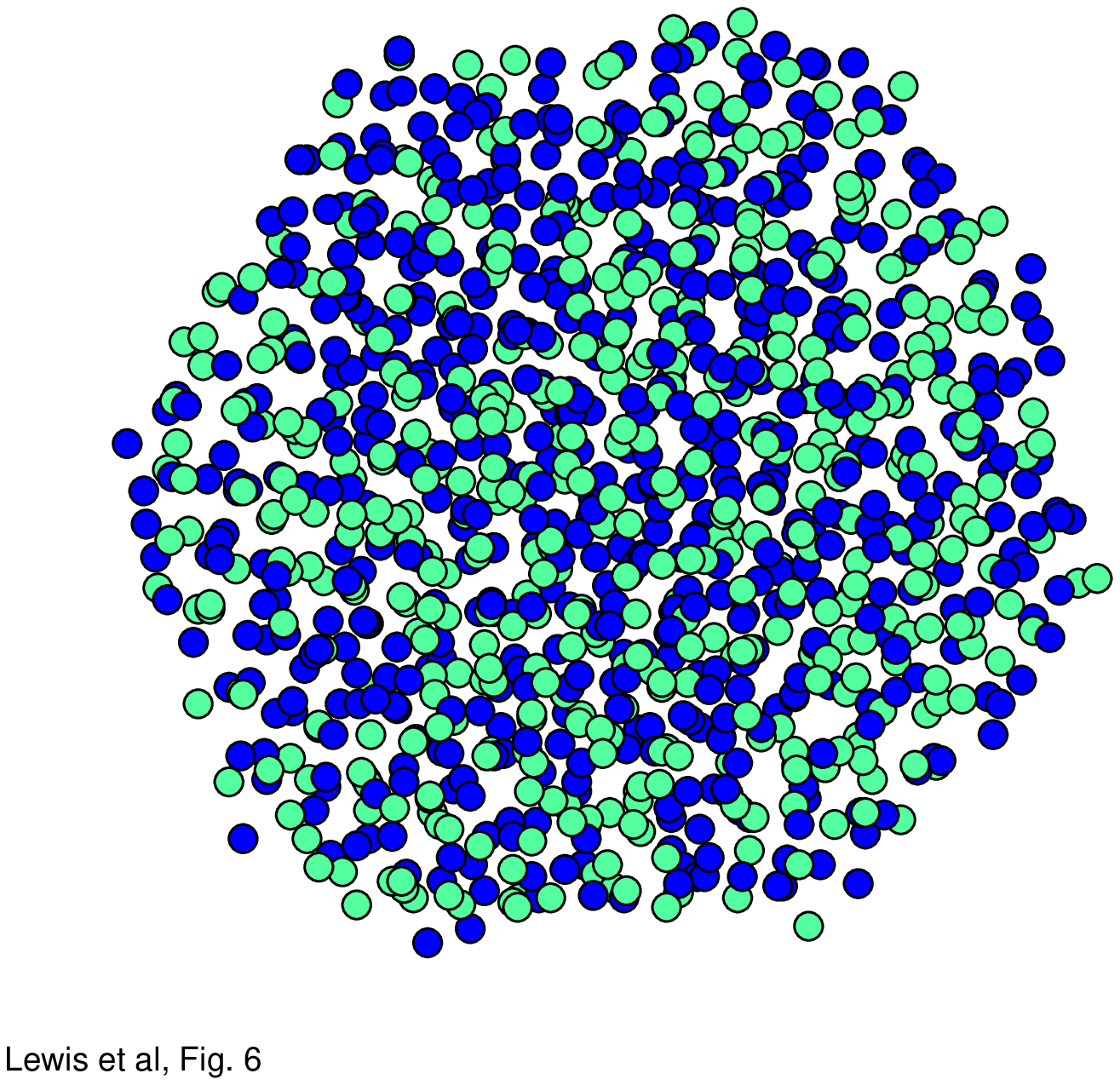}
\vspace*{-5cm}
\caption{
Structure of the 1055-atom cluster at 800 K during the quench process (i.e.,
in the liquid phase).
\label{1055_c800}
}
\end{figure}

Even more striking, however, is the fact that {\em even in the molten state},
the instantaneous shape of the cluster deviates markedly from spherical. We
see an example of this in Fig.\ \ref{1055_c800}, again for the 1055-atom
cluster, at a temperature of about 800 K, i.e., 90 degrees {\em above}
freezing. The surface of the cluster exhibits flattened regions, which
evidently are reminiscent of the facets that form on the surface of solid,
crystalline clusters. It is not clear whether this anisotropic shape reflects
some transient crystalline order in the vicinity of the surface, i.e., some
precursor fluctuations of the nearby transition, or wether it is related to
the deformation modes (breathing) of the liquid cluster, or a combination of
both. We have attempted to assess the existence of facets in a more
quantitative manner, but the quality of statistics, coupled to significant
thermal agitation at such high temperatures, is such that it is difficult to
draw unambiguous conclusions.

In any case, the fact that both the hot solid and the liquid (at least close
to the transition temperature) display anisotropic cluster shapes is an
important result. It will have important consequences for the coalescence of
clusters. Indeed, diffusion on a facet is very different from diffusion in a
(quasi)-spherical liquid overlayer. Far away from the edges of the facet,
barriers for diffusion are very similar to those found in a flat, infinite,
liquid overlayer, i.e., atoms do not ``feel'' the curvature of the cluster.
At the edges between two facets, further, the barriers can also be very
different and oppose diffusion.\cite{manninen} Thus, quite generally,
diffusion on a facetted cluster is expected to be significantly slower than
on a spherical cluster.

Upon cooling, as discussed earlier, a freezing transition takes place. In
Fig.\ \ref{nofr} we compare the distribution $N(r)$ for the initial cluster
at 703 K with that of the re-solidified cluster at (approximately) the same
temperature. It is clear, from this quantity, that the two clusters have
quite similar structures except in their outer shells, because of the
presence, in the re-solidified cluster, of extensive facets [such as those
seen in Fig.\ \ref{1055_800}(b)].

\subsection{Melting curve}

\begin{figure}
\vspace*{1cm}
\epsfxsize=7cm
\epsfbox{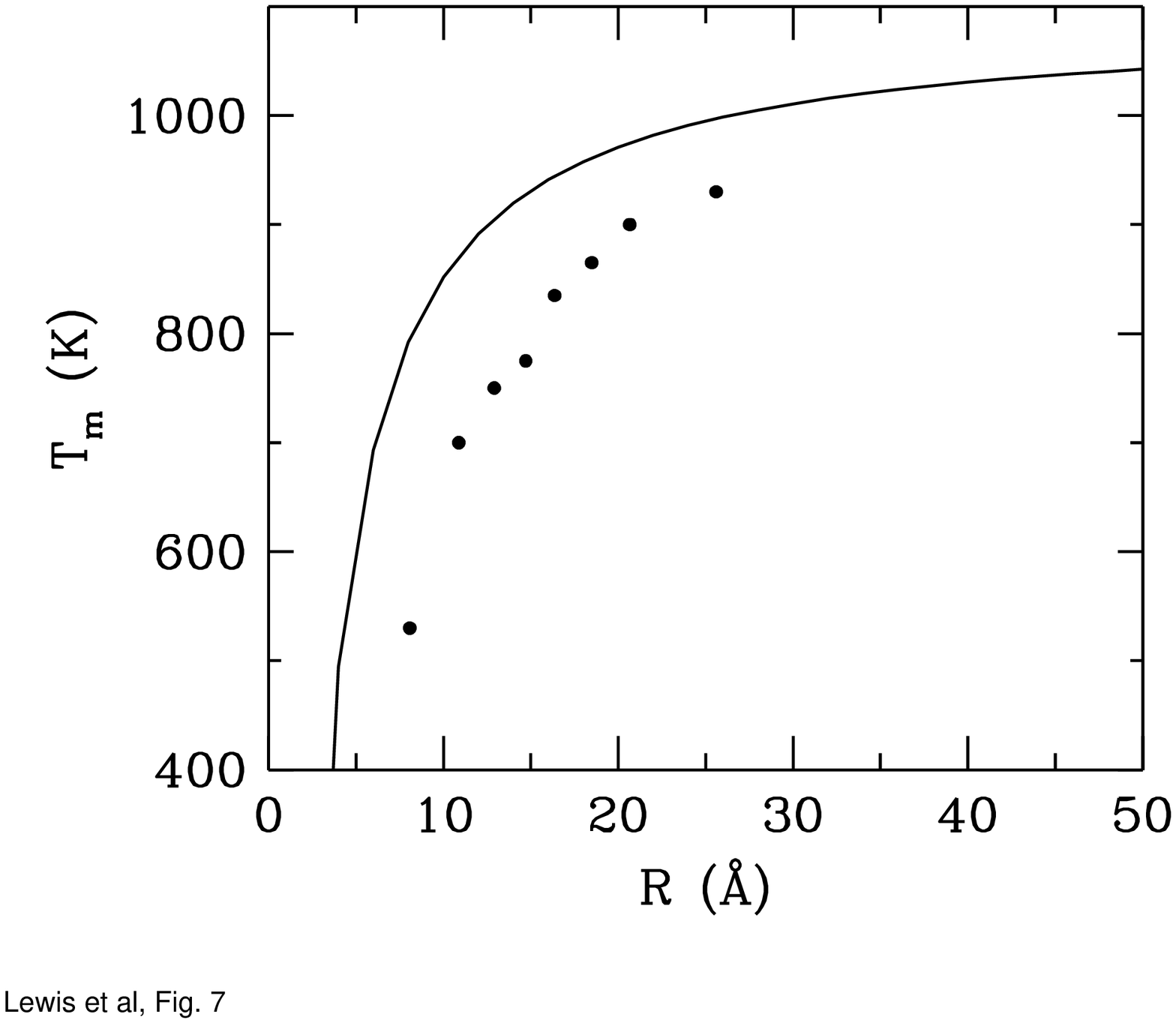}
\vspace{-1cm}
\caption{
Melting temperature as a function of cluster radius. The points indicate the
results of the simulations, while the solid line shows the predictions of the
simple thermodynamic model discussed in the text.
\label{T_m}
}
\end{figure}

The melting curve --- i.e., the variation of melting temperature with cluster
size (defined as $D=2R=2\sqrt{5/3}R_g$, where $R_g$ is the radius of gyration
of the cluster before melting), is displayed in Fig.\ \ref{T_m}. We also
show, on this graph, the results from a simple thermodynamic theory based
solely on the differences in surface energies of clusters of different sizes,
assuming the clusters are spherical:\cite{borel,stella}
   \begin{equation}
   {T(D)-T_{\infty} \over T_{\infty}} = -{4 \over \rho_S L D} \left[ \gamma_S -
   \gamma_L(\rho_S/\rho_L)^{2/3}\right],
   \label{fusionr}
   \end{equation}
where $T(D)$ is the melting temperature for a cluster of diameter $D$,
$T_{\infty}$ is the bulk melting temperature (1090 K for the present model),
$\rho_S$ the specific mass of the solid phase (19000 kg/m$^3$), $\rho_L$ the
specific mass of the liquid phase (17280 kg/m$^3$), $\gamma_S$ the surface
energy of the solid phase (0.9 J/m$^2$), $\gamma_L$ the surface energy of the
liquid phase (0.74 J/m$^2$), and $L$ the heat of fusion (53800 J/kg). Note
that, for consistency, all numerical values refer to the results from EAM
simulations with parameters appropriate to gold, rather than to experimental
data for the real material. We see that, despite the simplicity of the model,
the agreement is reasonable, comparable in fact to that with experimental
data;\cite{borel} just as is observed in experiment, Eq.\ \ref{fusionr}
systematically underestimates the deviation of the melting temperature with
respect to the bulk.

\section{Coalescence}\label{coalesc}

As a first step towards understanding the aggregation of clusters diffusing
on surfaces, and the subsequent pattern formation,\cite{PJLB} we have studied
the coalescence of pairs of free-standing (non-supported) clusters. As
discussed in the Introduction, this would correspond to the coalescence of
clusters on a surface with which they are only loosely bound (e.g., gold on
graphite). Three different cases were considered: coalescence of a liquid
cluster with another liquid, of a liquid and a solid, and of two solids. The
sintering of two single-crystal Cu nanoparticles was examined in Ref.\
\onlinecite{averback}. It is possible to study the coalescence of clusters in
various thermodynamic states because, as we have just seen, the melting
temperature depends on cluster size. Thus, at a given temperature, the state
of a cluster can be selected simply by selecting its size.

Here, we chose to investigate coalescence at a temperature of 800 K, which is
roughly in the middle of the range of melting temperatures for the single
clusters reported above (cf.\ Fig.\ \ref{T_m}) and therefore allows many
possible situations to be examined. We note that once coalescence is through
(or partly through), the melting temperature of the resulting cluster will be
different from that of the original clusters. Thus, for instance, two small
liquid clusters will result in a larger cluster which is also a liquid, while
two larger liquids would coalesce into a solid. (We have not studied the
latter situation).

In order to simulate the process, we use, as starting point configuration,
two fully-equilibrated clusters from the runs described in Section
\ref{melt_freeze}, i.e., retain, for each cluster, both the positions and the
velocities. The two clusters are placed in contact with one another (along
the $z$ axis), i.e., at a distance of approach roughly equal to the
nearest-neighbour distance for gold (2.89 \AA). We also ensure that the
initial angular momentum vanishes. For like clusters, we use configurations
from different points in time, and rotate them with respect to one another by
an angle which is chosen at random; indeed, it is expected that coalescence
will proceed differently if two facets are in contact than if a facet of one
cluster is in contact with a vertex of the other.

\subsection{Liquid-liquid}

\begin{figure}
\vspace*{1cm}
\epsfxsize=7cm
\epsfbox{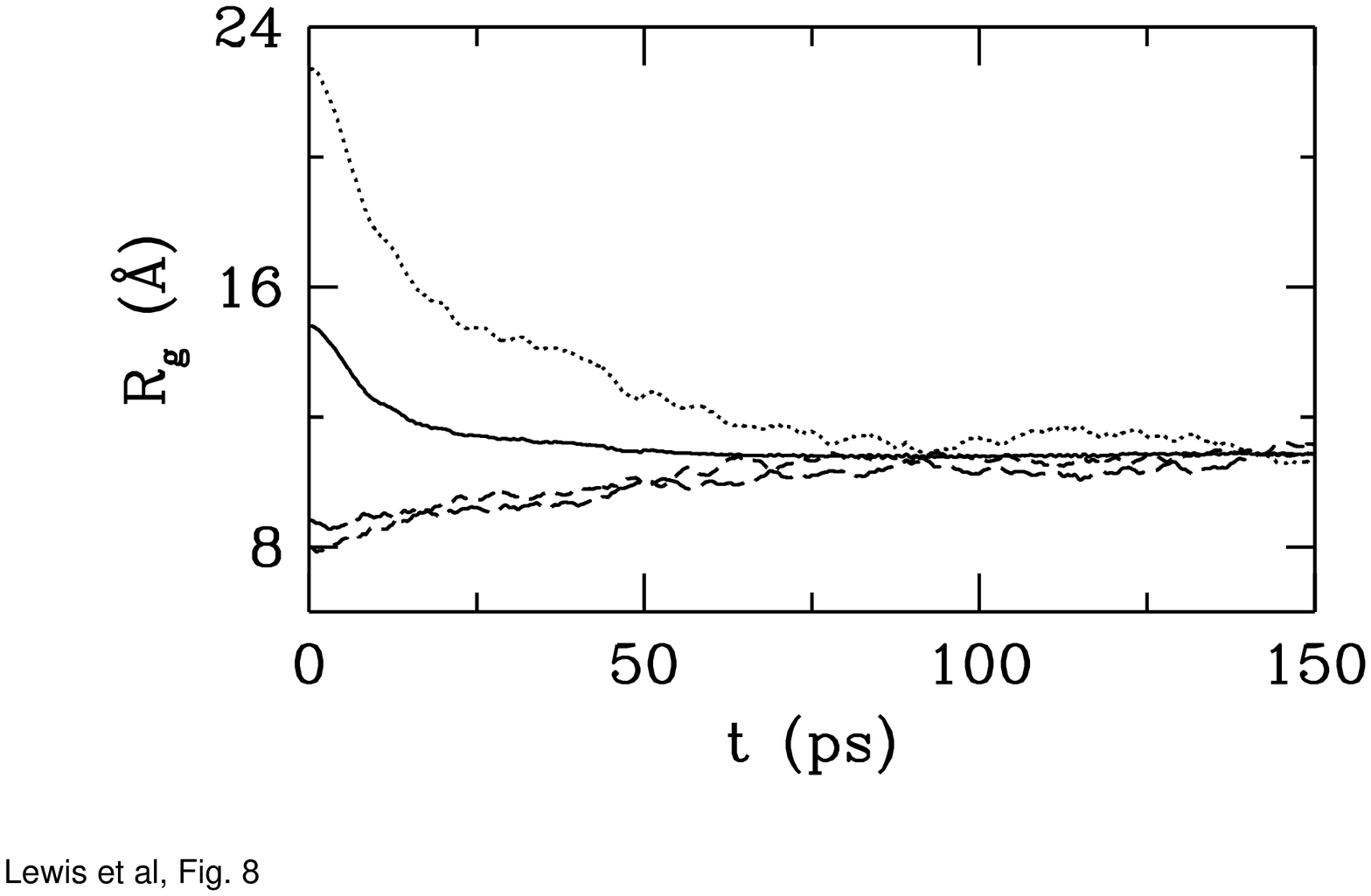}
\vspace*{-2cm}
\caption{
Radii of gyration for the coalescence of two 321-atom liquid clusters. The
dashed lines correspond to the $x$ and $y$ components, while the dotted line
is along $z$; the full line is the overall radius of gyration.
\label{321_321_R}
}
\end{figure}

\begin{figure}
\epsfxsize=10cm
\epsfbox{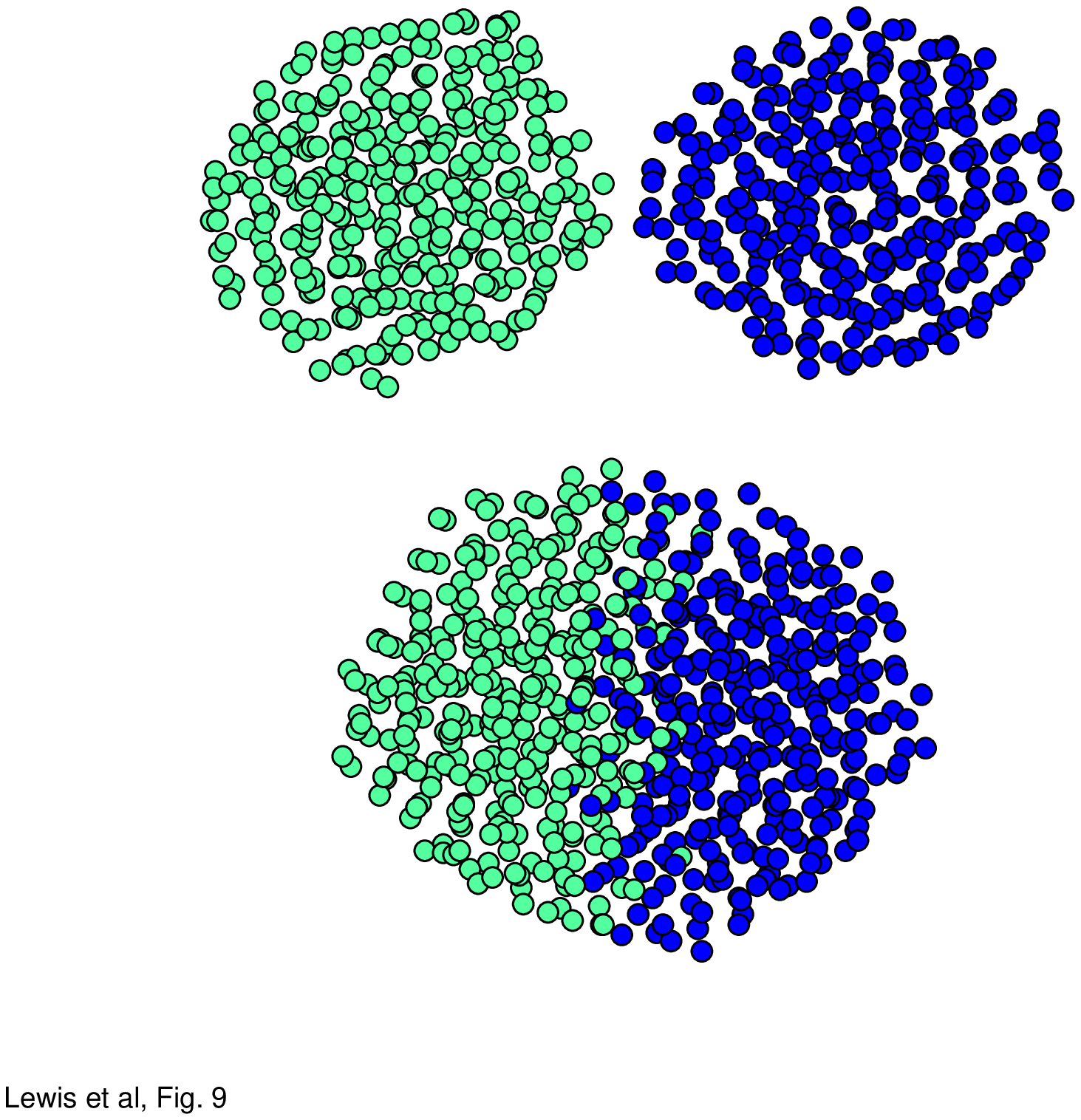}
\vspace*{-4cm}
\caption{
Coalescence of two 321-atom liquid clusters. Top: initial configuration;
bottom: after 75 ps.
\label{321_321_cfg}
}
\end{figure}

We present, first, results for two small liquid clusters --- 321 + 321 atoms.
From Fig.\ \ref{T_m}, we know that the resulting cluster will also be in a
liquid state. We show in Fig.\ \ref{321_321_R} the evolution in time of the
radii of gyration of the coalescing clusters, $R_{g\alpha}$ and $R_g$, where
$\alpha=x,y,z$, $R_g=\sqrt{R_{gx}^2+R_{gy}^2+R_{gz}^2}$, and
$R_{g\alpha}=(1/N)\sum_{i=1}^{N} (\alpha_i^2-\alpha_{cm}^2)$; $\alpha_{cm}$
is the $\alpha$-coordinate of the center of mass of the cluster, and $N$ is
the total number of particles. It is clear from this figure that the two
small clusters rapidly coalesce into a single, essentially spherical,
cluster: all three radii of gyration converge to the overall average on a
timescale of about 75 ps. A ball model of the system in its initial
configuration and at 75 ps is presented in Fig.\ \ref{321_321_cfg}. We have
colored the two initial clusters differently so as to better visualise the
process. It is interesting to note that coalescence into a spherical cluster
proceeds by the deformation of the two clusters in such a way as to optimize
the contact surface, i.e., {\em without} interdiffusion of one cluster into
--- or onto --- the other. Thus, the coalescence of two liquid clusters is
essentially a collective phenomenon, involving hydrodynamic flow driven by
surface tension forces. Of course, on longer timescales, since the cluster is
liquid, diffusion takes over and results in the mixing of the two initial
clusters.

\subsection{Liquid-solid}

\begin{figure}
\vspace*{1cm}
\epsfxsize=7cm
\epsfbox{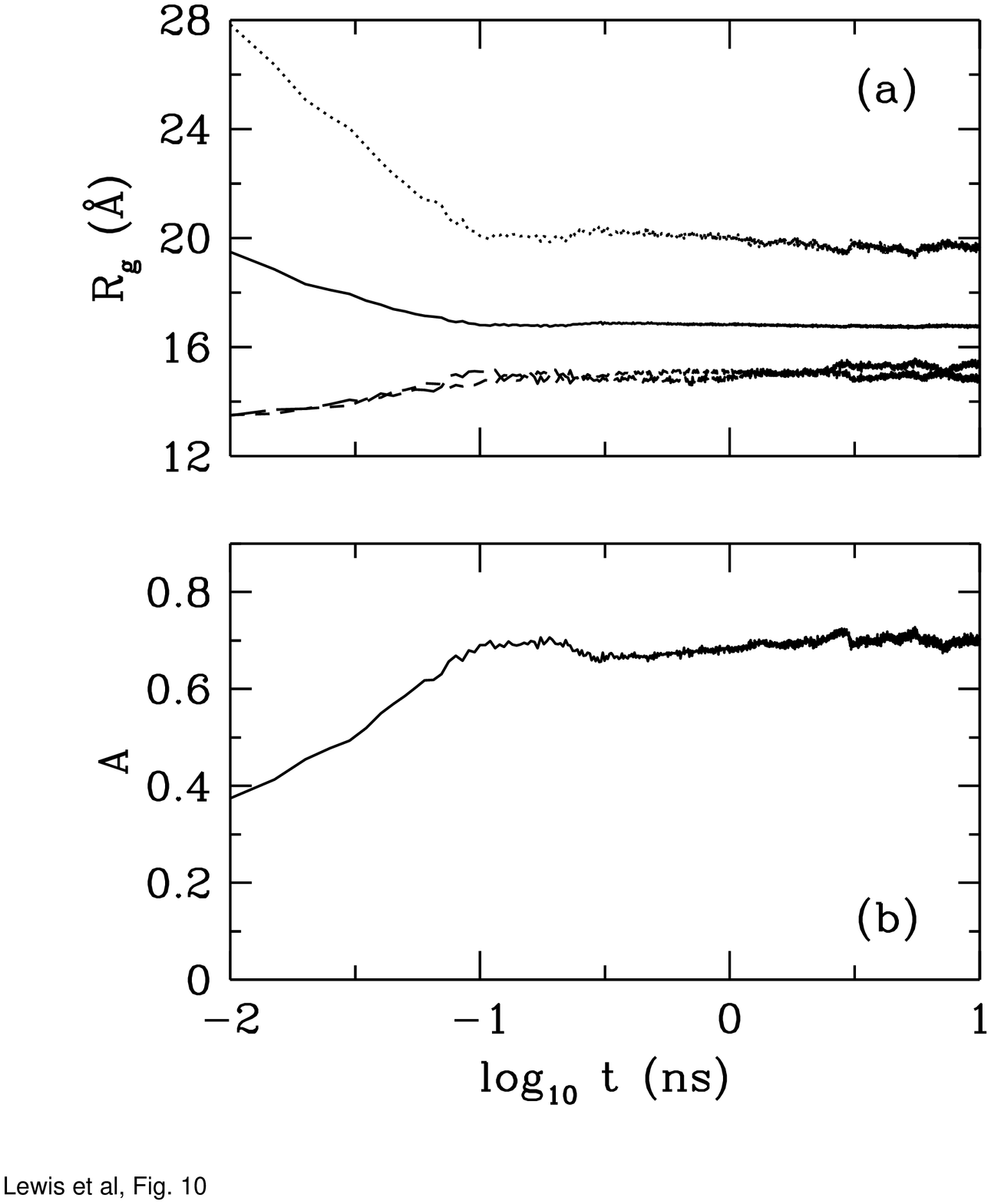}
\vspace*{1cm}
\caption{
(a) Radii of gyration and (b) aspect ratio, as defined in the text, versus
time, for the coalescence of a 767-atom liquid cluster with a 1505-atom solid
cluster.
\label{767_1505_R_M}
}
\end{figure}

\begin{figure}
\epsfxsize=10cm
\epsfbox{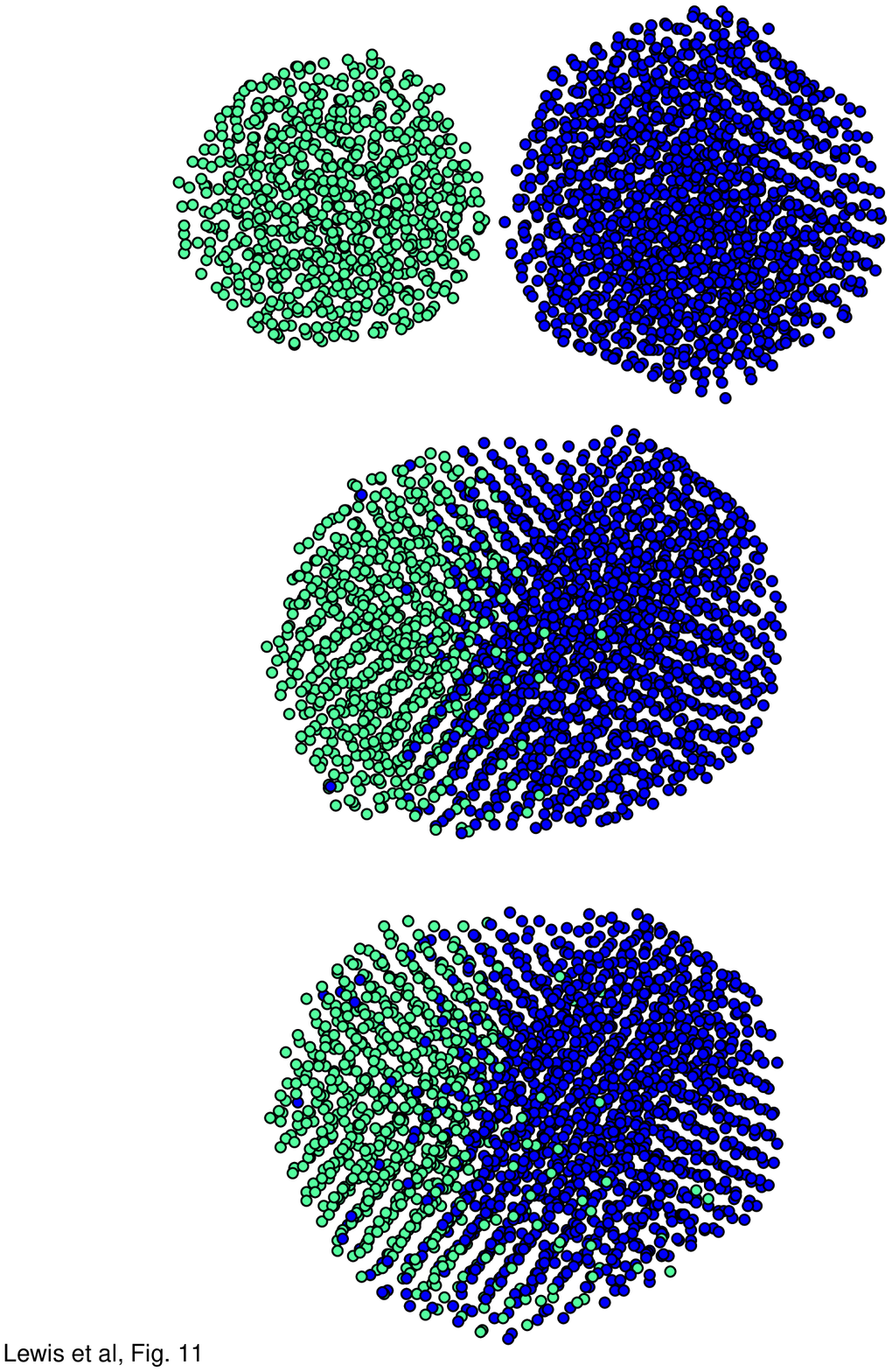}
\vspace*{-1cm}
\caption{
Configuration of the 767+1505 liquid-solid system at three different times:
0, 1 ns, and 10 ns (top to bottom).
\label{767_1505_cfg}
}
\end{figure}

We now examine the coalescence of a liquid cluster with a solid one taking,
as an illustration, the case 767 + 1505. The process in this case is expected
to be much slower than 321 + 321 and the dynamics, therefore, was followed
over a much longer time span of 10 ns. We plot in Fig.\ \ref{767_1505_R_M}(a)
the radii of gyration for this system. This reveals that coalescence proceeds
in two stages: First, maximizing the contact surface (against overall
volume), an extremely rapid approach of the two clusters is observed, taking
place on a timescale of about 100 ps. This, in fact, corresponds quite
closely to the time for coalescence of two liquid clusters, as we have seen
above. The cluster, at this point, is far from spherical but is nevertheless
smooth, possessing a facetted ovoidal shape. This can be seen in Fig.\
\ref{767_1505_cfg}, where we show the state of the cluster at a time of 1 ns
after the beginning of the coalescence. The initial stage of coalescence
(sintering) has been studied in some detail in Ref.\ \onlinecite{averback}
for two solid Cu clusters.

This rapid approach is followed by an extremely slow ``sphericization'' of
the system, driven by surface diffusion. This can in fact be inferred from
Fig.\ \ref{767_1505_R_M}(a). However, because the system can reorient in
space, the radii of gyration, which are defined with respect to a fixed
reference frame, do not provide a reliable characterization of the shape of
the cluster. In order to circumvent this difficulty, it is best to consider,
instead, the three principal moments of inertia; we plot, in Fig.\
\ref{767_1505_R_M}(b), the ratio of the smallest to the largest as a function
of time. This is in effect a measure of the ``aspect ratio'' of the
coalescing cluster; it should be noted that, by definition, this quantity is
always less than unity. Also, there are necessarily fluctuations in it that
are related to the modes of vibration of the cluster. Fig.\
\ref{767_1505_R_M}(b) reveals that, indeed, {\em modulo} some fluctuations,
the cluster exhibits a tendency to adopt a more spherical shape.

The timescale for the slow sphericization process is difficult to estimate
from Fig.\ \ref{767_1505_R_M}, but it would appear to be of the order of a
few hundred ns or more. In any case, this number is very substantially larger
than one would expect on the basis of phenomenological theories of the
coalescence of two soft spheres.\cite{nichols} Indeed, macroscopic theories
of sintering via surface diffusion\cite{nichols} predict a coalescence time
$\tau_c = k_BT R^4 /(C D_s \gamma a^4)$, where $D_s$ is the surface diffusion
constant, $a$ the atomic size, $\gamma$ the surface energy, $R$ the initial
cluster radius, and $C$ a numerical constant ($C=25$ according to Ref.\
\onlinecite{nichols}); taking $D_s \sim 5 \ 10^{-10}m^2s^{-1}$ (see Fig.\
\ref{msd}), $R=30$ \AA, $\gamma \approx 1 J m^{-2}$, and $a=3$ \AA, this
yields a coalescence time $\tau_c$ of the order of 40 ns.

\begin{figure}
\vspace*{1cm}
\epsfxsize=7cm
\epsfbox{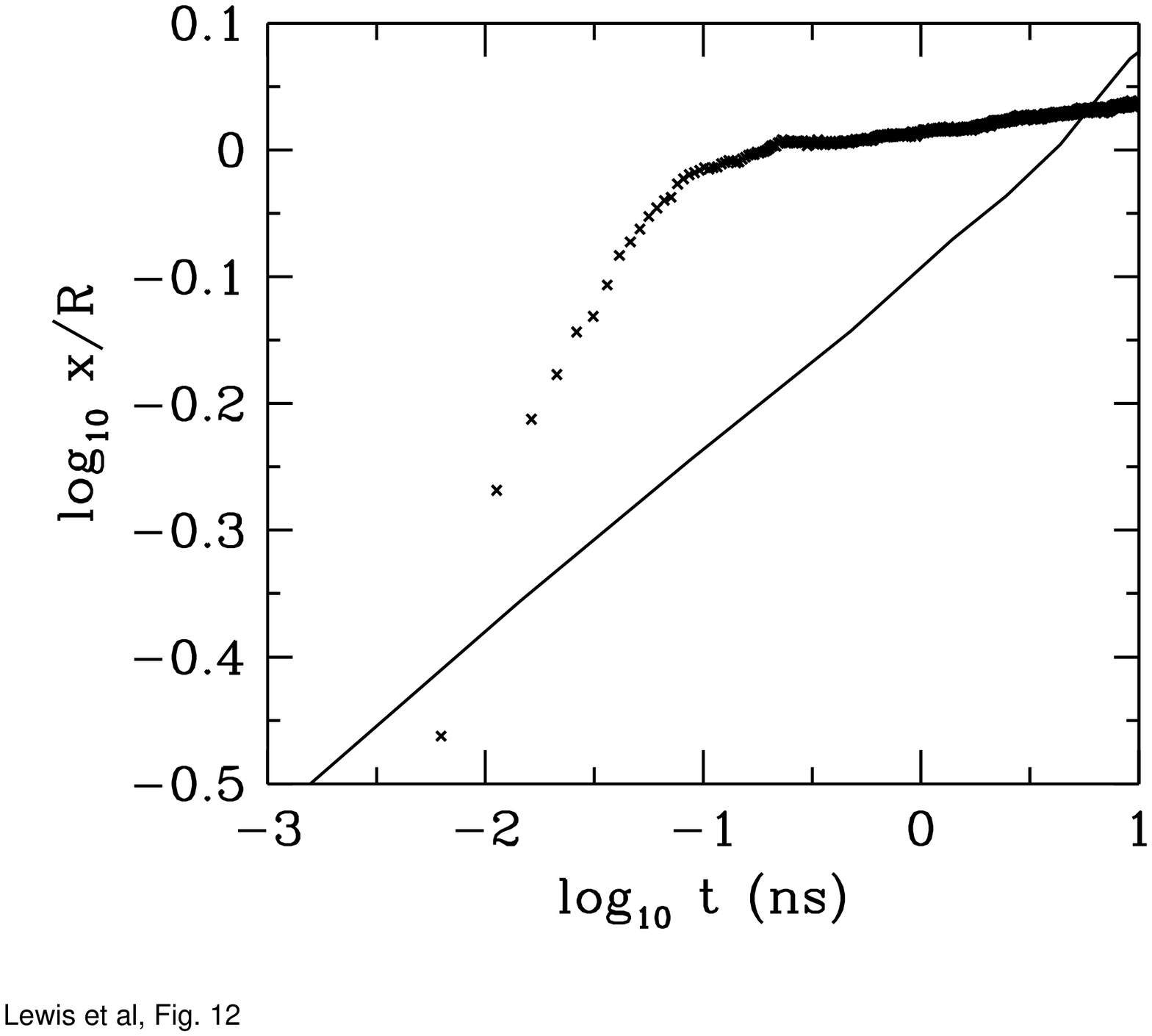}
\vspace{-1cm}
\caption{
Evolution in time of the ratio of the neck radius, $x$, to the cluster
radius, $R$. The full line represents the numerical solution obtained by
Nichols \protect\cite{nichols} while the crosses are the results of the
present simulations.
\label{neck}
}
\end{figure}

The same theories, in addition, make definite predictions on the evolution of
the shape of the system with time. In particular, in the tangent-sphere
model, the evolution of the ratio $x/R$, where $x$ is the radius of the
interfacial neck, computed numerically,\cite{nichols} is found to vary as
$x/R \sim (t/\tau_c)^{1/6}$ for values of $x/R$ smaller than the limiting
value $2^{1/3}$. In Fig.\ \ref{neck}, we compare the prediction of this
simple model (full line) with the results of the present simulations
(averaged over several different runs, including solid-solid coalescence ---
see below). There is evidently no possible agreement between model and
simulations. While the model predicts a uniform behaviour over a wide range
of timescales, we observe a very rapid growth at short times followed by an
extremely slow increase at long times. The rapid changes we see at short
times are due to elastic and plastic deformations not taken into account in
the numerical model; at long times, on the other hand, the presence of facets
slows down the diffusion, while the model assumes the cluster to be perfectly
spherical. As already noted above, our results are clearly not compatible
with a coalescence time $\tau_c$ of 40 ns; the actual value is evidently much
larger, by at least one or two orders of magnitude.

The much longer coalescence time we observe, again, is a consequence of the
presence of facets on the initial clusters, which persist (and rearrange)
during coalescence. The facets can be seen in the initial and intermediate
configurations of the system in Fig.\ \ref{767_1505_cfg}; the final
configuration of Fig.\ \ref{767_1505_cfg} shows that the cluster is more
spherical (at least from this viewpoint), and that new facets are forming. As
mentioned earlier, facets seriously limit the rate of diffusion that is
necessary to sphericize the cluster. Since a facet is flat, an atom diffusing
on it does not ``feel'' the curvature of the cluster and therefore behaves as
if it was on a flat, infinite surface --- except when approaching edges. We
have not examined this in detail for the case of gold, but in the case of
aluminum, the diffusion barriers at edges are found to be often larger than
on surfaces; for Al, e.g., diffusion from (100) to other facets is {\em not}
observed until temperatures very close to melting,\cite{manninen} and is
therefore not a favourable process. That diffusion is slow on our clusters
can in fact be seen from Fig.\ \ref{767_1505_cfg}: even after 10 ns, at a
temperature which is only about 200 degrees below melting for a cluster of
this size, only very few atoms have managed to diffuse a significant distance
away from the contact region.

We have not analyzed in detail the structure (and its evolution in time) of
the cluster. However, it is clear from Fig.\ \ref{767_1505_cfg} that already
at 1 ns, the system is completely solid, i.e., the initially-liquid cluster
has solidified upon coalescing with the larger solid cluster. Visual
inspection of the configurations indicates that the timescale for
solidification is roughly the same as that for the initial approach --- about
100 ps.

\subsection{Solid-solid}

\begin{figure}
\vspace*{1cm}
\epsfxsize=7cm
\epsfbox{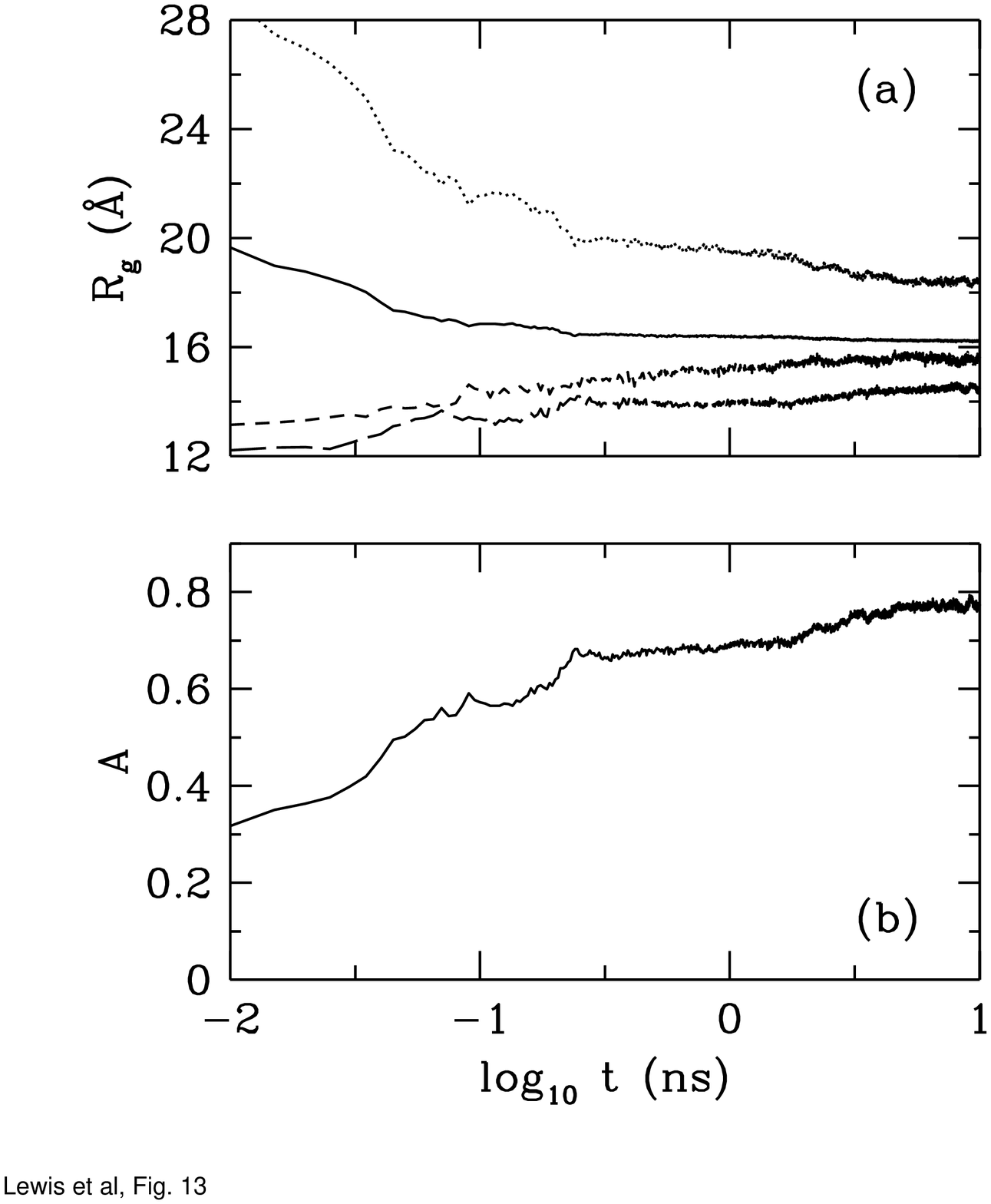}
\vspace*{1cm}
\caption{
(a) Radii of gyration and (b) aspect ratio, versus time, for the coalescence
of two 1055-atom solid clusters.
\label{1055_1055_R_M}
}
\vspace{3cm}
\end{figure}

\begin{figure}
\epsfxsize=10cm
\epsfbox{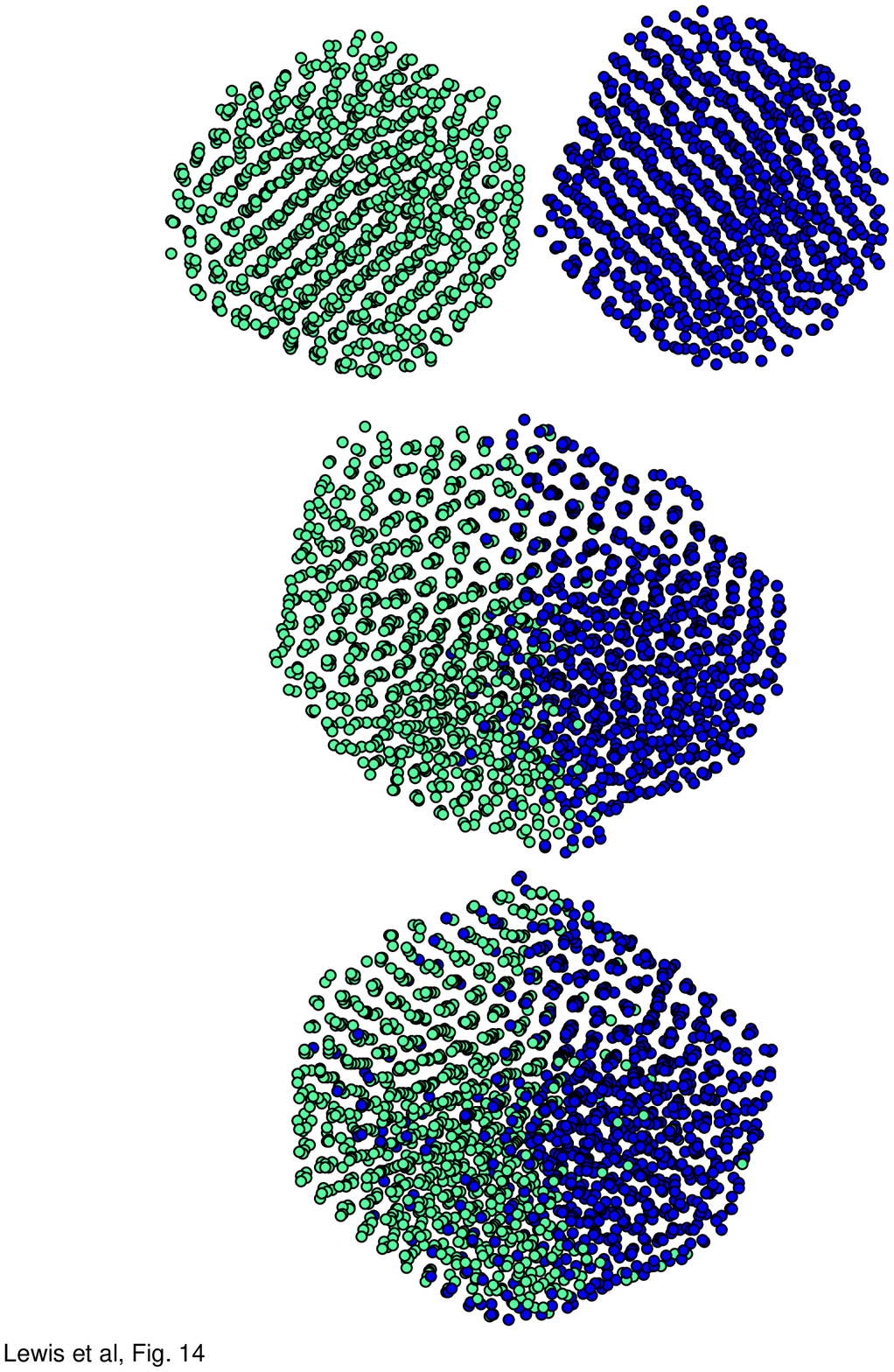}
\vspace*{-1cm}
\caption{
Configuration of the 1055-1055 solid-solid system at three different times:
0, 1 ns, and 10 ns (top to bottom).
\label{1055_1055_cfg}
}
\end{figure}

Finally, we have also examined the coalescence of two solid, 1055-atom
clusters. We give in Figs.\ \ref{1055_1055_R_M}(a) and (b) the evolution in
time of the radii of gyration and of the aspect ratio, as defined above,
respectively. In Fig.\ \ref{1055_1055_cfg} we show the initial state of the
system as well as at times of 1 and 10 ns.

The behaviour of the system in this case is analogous to that found for the
liquid-solid case (passed the initial approach). One apparent difference,
however, is that coalescence seems to proceed {\em faster} here than it did
for the liquid-solid case; this can be seen upon comparing the aspect ratios,
Figs.\ \ref{1055_1055_R_M}(b) and \ref{767_1505_R_M}(b). Note that since the
present system is larger, the diffusion should be relatively slower, since
the distance in temperature to the melting point is larger, i.e.,
sphericization should proceed more {\em slowly}.

Of course, we cannot draw general conclusions based on these two particular
examples, and it is certainly the case that the coalescence of a liquid and a
solid proceeds faster, in general, than the coalescence of two solids. (We
have observed such cases). The particular behaviour we observe here is likely
related to the internal structure of the cluster: Judging from Fig.\
\ref{767_1505_cfg} we see that the internal structure of the 767-1505-atom
cluster is complex and perhaps ``grainy'', i.e., consisting of grains or
domains. Thus, there are high-energy extended defects, or grain boundaries,
that prevent crystallization into a single domain from taking place. In
contrast, for the 1055-1055-atom system, we seem to have more of a
single-domain structure. Thus, the coalescence of two solid clusters is
expected to depend strongly on the ``initial conditions'', i.e., relative
orientations, while this should not be the case for the coalescence of a
liquid and a solid, though misfits might develop.

\section{Concluding remarks}\label{concl}

In this paper, we have presented a MD-EAM study of some dynamic and
thermodynamic properties of unsupported gold nanoparticles. With regards to
thermodynamics, we have focused on the melting transition of small (less than
3 nm) clusters. Our results are consistent with those reported by Tosatti et
al.,\cite{tosatti} and Yu and Duxbury,\cite{duxbury} using a many-body
``glue'' Hamiltonian. In particular, we observe that melting proceeds from
the surface inwards, i.e., there exists a dynamical ``premelting'' of the
outer layers signalling the approach of the melting point. A noteworthy
result of our simulations is the evidence for a large (a few hundred Kelvins)
melting hysteresis for the clusters, in qualitative agreement with
experiments on lead.\cite{kofman}

\begin{figure}
\vspace*{1cm}
\epsfxsize=7cm
\epsfbox{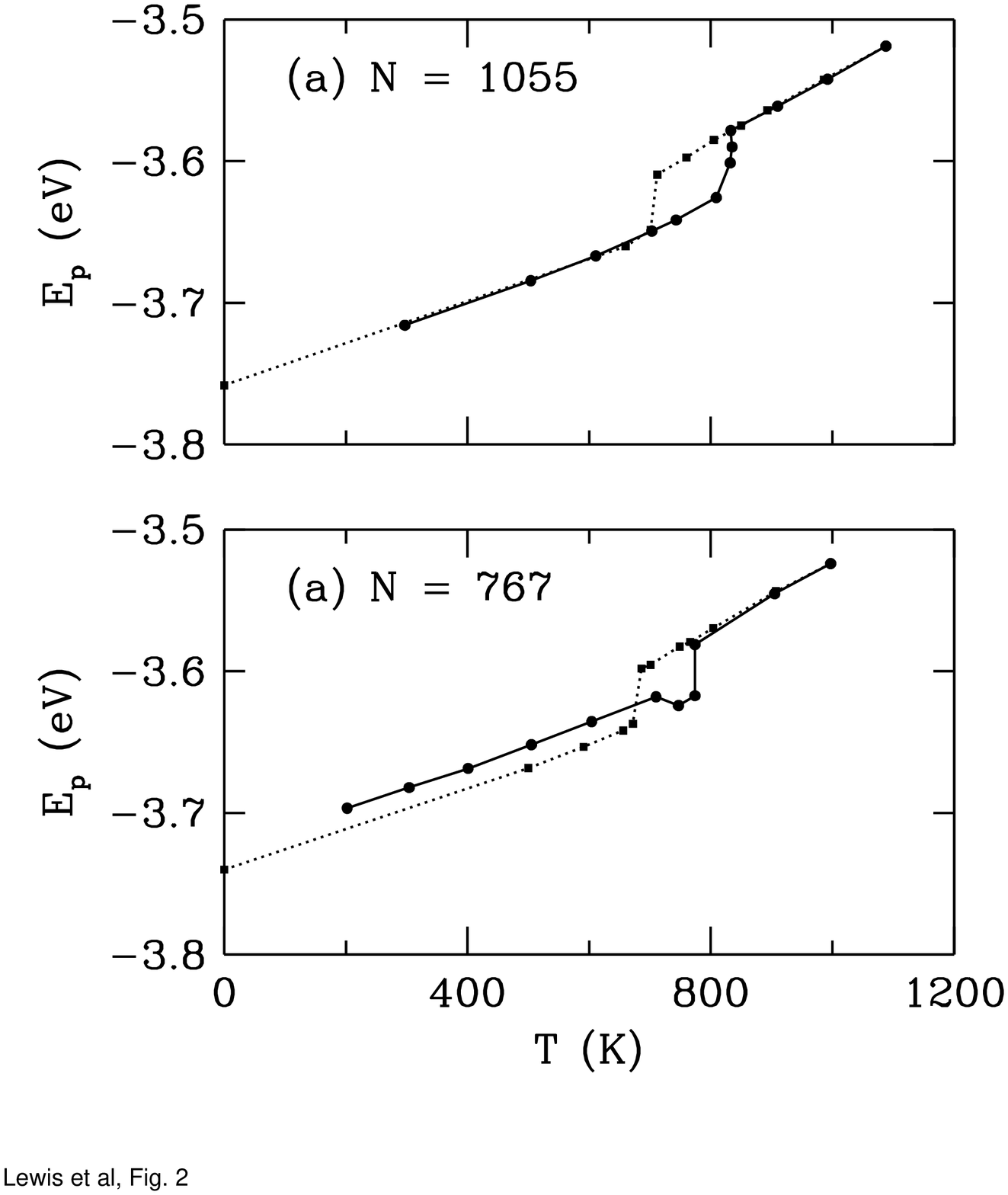}
\vspace*{1cm}
\caption{
Potential energy (per atom) as a function of temperature for (a) the
1055-atom cluster and (b) the 767-atom cluster; full and dotted curves
correspond to heating and cooling, respectively.
\label{E_vs_T}
}
\end{figure}

As in earlier experimental or numerical work, the analysis of the
melting-freezing cycle raises questions on the applicability of macroscopic
concepts --- such as crystal, liquid, and surface tension --- to clusters
consisting of more than a hundred atoms or so. We find, in particular, that
the deviation of the melting temperature from the bulk value is reasonably
well described in terms of such concepts. Clearly, however, the clusters
exhibit polymorphism and structural transitions that are not taken into
account in this microscopic approach (see Fig.\ \ref{E_vs_T}), but these
aspects do not seem to be essential when the number of atoms exceeds a few
hundred.

This situation for the melting-freezing transition is in sharp contrast with
our findings for cluster coalescence. Here, the macroscopic theories of
sintering via surface diffusion completely fail, both qualitatively and
quantitatively, to describe the coalescing behaviour of two small clusters.
We attribute this failure to the fact that the nanocrystals are facetted,
while the sintering theories are formulated for macroscopically smooth
crystallites. In order to attain the spherical equilibrium shape that can be
expected from thermodynamic considerations, a cluster has to reduce the
number of its facets. Such a process, in turn, requires collective
rearrangement of the atoms, with correspondingly high energy barriers. To our
knowledge, the difficult problem of the approach to equilibrium of a facetted
crystal has not, up to now, been investigated theoretically.

The fact that the coalescence process is ``slow'', or slower than expected,
has important consequences for the morphology of cluster-assembled materials.
Indeed, the type of morphology --- compact or ramified --- depends critically
on the ratio between the coalescence time and the time it takes for a new
cluster to join an existing group.\cite{PJLB} If this ratio is larger than
one, ramified objects are expected to form, as observed, for instance, in
Ref.\ \onlinecite{PJLB}. In the opposite case, compact objects will result,
and the material will be unable to retain memory of the initial ``building
blocks'' from which it was formed, thus leading to a smooth, uniform
structure. It is therefore important to study further coalescence in order to
characterize such effects in more detail.

\acknowledgements

It is a pleasure to thank R. Kofman (Universit\'e de Nice) and P. M\'elinon
(UCBL) for many useful discussions. LJL is thankful to Profs.\ Serughetti and
Barrat for the invitation at the DPM/UCBL where most of the work presented
here was carried out, as well as the personnel of the laboratory for their
kind hospitality; financial support from the CNRS is also gratefully
acknowledged. This work was also supported by the P\^ole Scientifique de
Mod\'elisation Num\'erique at \'ENS-Lyon, the Natural Sciences and Engineering
Research Council of Canada, and the ``Fonds pour la formation de chercheurs
et l'aide \`a la recherche'' of the Province of Qu\'ebec.



\begin{center}
\begin{table}
\caption{
Melting temperature of the various clusters considered in the present study
as a function of their radius just before melting, $R$ (given, in terms of
the radius of gyration $R_g$ by $R=\protect\sqrt{5/3}R_g$).
}
\label{clusters}
\begin{tabular}{lcc}
 $N$ & $R$ (\AA) & $T_m$ \\ \hline
 135 &   8.08    & 530 \\
 321 &  10.89    & 700 \\
 531 &  12.92    & 750 \\
 767 &  14.72    & 775 \\
1055 &  16.37    & 835 \\
1505 &  18.50    & 865 \\
2093 &  20.68    & 900 \\
3997 &  25.62    & 930 \\
\end{tabular}
\end{table}
\end{center}

\newpage

\newpage

\end{document}